\documentclass[useAMS,usenatbib]{mnras}
\usepackage{graphicx}\usepackage{epsfig}\usepackage{epsf}
\usepackage{amsmath}
\usepackage{amssymb}
\usepackage{stfloats}
\usepackage{lipsum}
\usepackage{pgf}
\usepackage{tikz} 
\usetikzlibrary{shapes,arrows,automata}
\usepackage{tree-dvips}
\usepackage{soul}
\usepackage{caption}

\title[SN\,2017hcc spectropolarimetry]{Record-breaking polarization from the interacting superluminous supernova 2017hcc}

\author[J. C. Mauerhan et al.]{Jon C. Mauerhan,$^{1}$\thanks{E-mail: jon.c.mauerhan@aero.org}
Nathan Smith,$^{2}$
G. Grant Williams,$^{2,3}$
Paul S. Smith,$^{2}$
Alexei V.\ Filippenko,$^{4}$
\newauthor Christopher Bilinski,$^{2}$ WeiKang Zheng,$^{4,5}$, Thomas G. Brink,$^{4,6}$, Jennifer L. Hoffman,$^{7}$ \newauthor Douglas C. Leonard,$^{8}$ Peter Milne,$^{2}$ Benjamin Jeffers,$^{4}$ Shaunak Modak,$^{4, 9}$ Samantha Stegman,$^{4}$ \newauthor Keto D. Zhang$^{4,9}$
\\
$^{1}$The Aerospace Corporation, 2310 E. El Segundo Blvd., El Segundo, CA 90245, USA\\
$^{2}$Steward Observatory, University of Arizona, 933 N. Cherry Ave., Tucson, AZ 85721, USA \\
$^{3}$MMT Observatory, Tucson, AZ 85721-0065, USA \\
$^{4}$Department of Astronomy, University of California, Berkeley, CA 94720-3411, USA \\
$^{5}$Eustace Specialist in Astronomy \\
$^{6}$Wood Specialist in Astronomy \\
$^{7}$ Department of Physics and Astronomy, University of Denver, 2112 East Wesley Avenue, Denver, CO 80208, USA\\
$^{8}$ Department of Astronomy, San Diego State University, PA-210, 5500 Campanile Drive, San Diego, CA 92182-1221, USA\\
$^{9}$ Department of Astrophysical Sciences, Princeton University, 4 Ivy Lane, Princeton, NJ 08544, USA\\
$^{10}$ IPAC, Mail Code 100-22 Caltech, 1200 E. California Blvd. Pasadena, CA 91125, USA\\
}
\pubyear{2023}
\begin{document}
\label{firstpage}
\pagerange{\pageref{firstpage}--\pageref{lastpage}}
\maketitle
\begin{abstract}
We present multiepoch spectropolarimetry of the superluminous interacting Type IIn supernova (SN) 2017hcc, covering 16 to 391\,days after explosion. In our first  epoch we measure continuum polarization as high as 6\%, making SN\,2017hcc the most intrinsically polarized SN ever reported.  During the first 29\,days of coverage when the polarization is strongest, the continuum polarization has a wavelength dependence that rises toward blue wavelengths, whereas the continuum polarization becomes wavelength independent by day 45. The polarization strength drops rapidly during the first month, even as the SN flux is still climbing to peak brightness. Nonetheless, record-high polarization is maintained until day 68, at which point the source polarization declines to 1.9\%, comparable to peak levels in previous well-studied SNe~IIn.  Thereafter the SN continues in polarization decline, while exhibiting only minor changes in position angle on the sky. The blue slope of the polarized continuum during the first month, accompanied by short-lived polarized flux for Balmer emission, suggests that an aspherical distribution of dust grains in pre-shock circumstellar material (CSM) is echoing the SN~IIn spectrum and strongly influencing the polarization, while the subsequent decline during the wavelength-independent phase appears broadly consistent with electron scattering near the SN/CSM interface. The persistence of the polarization position angle between these two phases suggests that the pre-existing CSM responsible for the dust scattering at early times is part of the same geometric structure as the electron-scattering region that dominates the polarization at later times. SN\,2017hcc appears to be yet another, but much more extreme, case of aspherical yet well-ordered CSM in Type IIn SNe, possibly resulting from pre-SN mass loss shaped by a binary progenitor system. 

\end{abstract}




\section{Introduction}
The namesake relatively narrow hydrogen lines of Type IIn supernovae (SNe~IIn) arise when moderately slow circumstellar material (CSM) around the progenitor is shocked or photoionized by the SN (see \citealt{smith17} for a review of interacting SNe).
Polarimetric observations of SNe~IIn commonly show high degrees of continuum polarization relative to other classes of SNe \citep{Mauerhan2014, Bilinski2018,Bilinski2021,Hoffman2008}. The most widely invoked hypothesis for the strong polarization is electron scattering at the interface of the SN ejecta and dense aspherical CSM, creating an optically-thick pseudophotosphere that traces the geometry of the CSM. 

In principle, scattering of SN photons off circumstellar dust grains could also  generate continuum polarization in SNe~IIn, but dust-induced polarization has not been considered in most previous studies of interacting SNe; see, however, Mauerhan et al. (2017b) and Bilinski (2021). If dust scattering contributes to the polarized flux, these reflected photons should mix with the electron-scattered photons to create complex spectropolarimetric evolution.  In this paper, we consider both sources of polarization and their implications for the relative geometries of the gas and dust distributions. 

Among the most well-studied polarized SNe was the 2012 explosion of SN\,2009ip \citep{Mauerhan2013}, which resulted in one of the most complete spectropolarimetric datasets ever obtained for an SN~IIn, and provided a striking example of multicomponent geometry in an SN environment. \citet{Mauerhan2014} demonstrated that the so-called 2012a outburst, during which broad emission lines from a fast SN outflow first emerged, was strongly polarized along a particular axis of symmetry. Roughly a month later, the 2012b event began when the SN blast began interacting with the CSM created by stellar outbursts from years prior  \citep{Smith2010}, and a strong jump in polarization was accompanied by a $\sim 90^{\circ}$ flip in the on-sky position angle of the polarized source. This behavior implies that the SN and CSM had orthogonal geometries, consistent with an intrinsically bipolar explosion that crashed into an equatorial or disk-like distribution of CSM.  Significant asymmetry or axisymmetry in the CSM of SNe~IIn has important implications for the mechanism of their eruptive pre-SN mass loss \citep{Smith2014,sa14}.  It has been suggested that pre-SN mass loss shaped by binary interactions between evolved stars at late stages of nuclear burning possibly ejects mass into the equatorial plane of the binary \citep{sa14}, setting the stage for the type of spectropolarimetric evolution observed in SN\,2009ip \citep{Mauerhan2014}. It is important to note that spectropolarimetric data obtained during the earliest phases of SN\,2009ip's 2012 event were essential in revealing the orthogonal SN/CSM geometry, made possible only because of the early discovery of the eruptive progenitor as a luminous blue variable \citep{Smith2010}. 

SN\,2017hcc was a relatively nearby interacting explosion (Type~IIn) that achieved superluminous status \citep{Prieto2017}, located in  an anonymous host galaxy at redshift $z=0.0168$. The explosion date adopted here is MJD 58027.9, estimated from the well sampled rise of the light curve  \citep{Prieto2017}. During the early evolution no X-rays or radio emission were detected from the source \citep{Chandra2017,Nayana2017}. As in the case of SN~2006gy, a slow rise to a superluminous optical peak - but with no X-ray emission - indicates that the early CSM interaction was very optically thick \citep{Smith07gy,SM07}.  At much later times on day 727 a faint X-ray source was detected, and after 1000\,days at radio wavelengths \citep{Chandra2022}; analysis indicates an extreme mass-loss episode in the decade prior to explosion, which created an dense inner envelope that the shock broke through during the initial days to months. \citet{Chandra2022} also noted an unexpectedly high ratio of infrared (IR) to X-ray luminosity, interpreted as possible evidence for an asymmetric circumstellar region. 

\citet{Smith2020} presented a detailed analysis of SN\,2017hcc, which included high-resolution spectroscopy.  They interpreted the evolving blueshifted emission-line profiles as the result of dust condensation in post-shock material, and noted interesting similarities with SN\,2010jl. \citet{Smith2020} explained the shape and evolution of narrow-line profiles in SN\,2017hcc by invoking CSM having bipolar morphology that we view from a mid-latitude, where early phases included strong CSM interaction with the pinched equatorial waist of the bipolar CSM shell. \citet{moran2023} also presented additional spectra of SN\,2017hcc, including IR spectroscopy, which they interpreted as evidence for pre-existing CSM dust.

\citet{Mauerhan2017a} reported an early epoch of spectropolarimetry exhibiting an integrated $V$-band polarization near 5\%, the strongest polarization ever reported for any SN. \citet{Kumar2019} subsequently published broad-band imaging polarimetry of this explosion with coverage starting at a later date, confirming a declining yet persistently strong polarization signature. Some of the spectropolarimetric data analyzed below in this paper also appear in the study by Bilinski et al. (2023, in prep.), where spectropolarimetry of SN2017hcc is analyzed as part of a larger sample of SNe IIn.  For consistency of method in analyzing that larger sample, some of Bilinski et al.'s adopted values (ISP, E(B-V), etc.) are slightly different than those used here, but the results are complimentary.  This paper presents our full spectropolarimetric dataset on SN\,2017hcc, including the earliest spectropolarimetry of the source, which exhibits even stronger polarization than previously reported. 

\begin{table*}
\caption{Integrated spectropolarimetry of SN\,2017hcc.} 
\renewcommand\tabcolsep{14pt}
\begin{tabular}[b]{@{}clrccc}
\hline
Epoch & UTC dates averaged& day range $^{\textrm{a}}$ &Tel./Instr.   & $P$(\%) &   $\theta$ (deg)  \\
 \hline
  \hline
1&Oct. 17 & $16$ &Kuiper/SPOL &  5.80 (0.04)  & 92.6 (0.1)    \\
2&Oct. 20--22 & $19-21$ &Kuiper/SPOL &  5.69 (0.02)  & 94.4 (0.1)    \\
3&Oct. 30 & $29$ &Lick/Kast &  4.79 (0.03)   & 96.4 (0.1)    \\

4&Nov. 15--16 & $45-46$ &Kuiper/SPOL  &  3.08 (0.02)   & 97.8 (0.1)    \\
5&Nov. 21--22 & $51-52$ &Kuiper/SPOL  &  2.57 (0.02)   & 99.2 (0.1)    \\

6&Dec. 8--9  & $68-69$ & Kuiper/SPOL&  1.91 (0.02)   & 102.5 (0.2)   \\
7&Dec. 12  &  72  &Lick/Kast &   1.59 (0.12)   & 107.5 (0.8)    \\

8&Dec. 15--16 & $75-76$ & Kuiper/SPOL&  1.83 (0.02)   & 102.4 (0.2)   \\

9&Dec. 21--22  &  81--82& MMT/SPOL &   1.65 (0.01)   & 104.1 (0.2)    \\

10&Jan. 12--16   &  103-107  &Kuiper/SPOL  & 1.60 (0.10)   & 100.0 (5.1)    \\
11&Jan. 19--20, 22--23  & 110-114& Bok/SPOL & 1.06 (0.03)   &  103.9 (1.8)   \\
12&Jul. 19 (2018)  &  291  &Lick/Kast &   <0.07 (0.08)   & --    \\
13&Oct. 28,29 Nov. 1 (2018) &391-393  & Bok/SPOL & $<$0.29  &  --   \\

\hline \\
\end{tabular}

$^a${Day range is with respect to the estimated explosion date JD~2,458,027.9. All integrated values of $P$ and $\theta$ are for the $V$ band (5050--5950\,{\AA}), except for the two latest 2018 epochs of relatively low S/N, which were integrated over the range 6400--6700\,{\AA} to provide upper limits on the continuum polarization.}
\end{table*}

\section{Observations}
Spectropolarimetry was obtained at three of the University of Arizona's observatories. The Kuiper 61\,in telescope at Mt. Lemmon, AZ, was used on 2017 Oct. 17--22, Nov. 15--22, Dec. 8--16, and 2018 Jan. 12--16 UTC with the CCD Imaging/Spectropolarimeter (SPOL;  \citep{Schmidt1992}. On Dec. 21 the 6.5\,m Multiple Mirror Telescope (MMT) on Mt. Hopkins was used with the same SPOL instrument. A very late epoch the following year was obtained using the 2.3\,m Bok telescope on Oct. 28, 29, and Nov. 1, also with the SPOL instrument. The University of California's Shane 3\,m reflector at Lick Observatory and Kast instrument \citep{Miller1988} were used on 2017 Oct. 30 and Dec. 12, as well as on 2018 July 19. Both SPOL and Kast are dual-beam spectropolarimeters that utilize a rotatable semiachromatic half-wave plate to modulate incident polarization and a Wollaston prism in the collimated beam to separate the two orthogonally polarized spectra onto CCD sensors. Additional details of these instruments can be found in the above references and in \citep{Mauerhan2014}. The instrument polarization angle and response to polarized light are calibrated using observations of polarized and unpolarized standard stars, respectively.  All spectral data were extracted and calibrated in the standard manner using generic IRAF routines and home-grown IDL functions. Spectropolarimetric analysis was also performed in IRAF and IDL following the methods described by \citep{Miller1988} and implemented by \citet{Leonard2001}. After all calibrations were applied, the data were corrected for the redshift of the host galaxy.

Multicolor photometry of SN\,2017hcc was obtained using the 0.76\,m Katzman Automatic Imaging Telescope \citep[KAIT;][]{Filippenko2001}) and the 1\,m Nickel telescope at Lick Observatory, with the first observation collected on MJD 58054.30441. The data are listed in the Appendix.

\section{Analysis \& Results}
Polarization is generally expressed as the quadratic sum of the $Q$ and $U$ Stokes parameters, $P= \sqrt{q^2 + u^2}$, and the position angle on the sky is given by $\theta_{sky}=1/2~\textrm{tan}^{-1}(u/q)$, while carefully taking into account the quadrants in the $Q$--$U$ plane where the inverse-tangent angle is located. Since $P$ is a positive-definite quantity, it is biased high in situations where the signal-to-noise ratio (S/N) is low.  It is thus customary to express the ``de-biased'' form of $P$, given by $P_{\rm db}=\sqrt{\lvert (q^2 + u^2) - (\sigma_{q}^2 + \sigma_{u}^2) \rvert}$, where $\sigma_q$ and $\sigma_u$ represent the uncertainties in the $Q$ and $U$ Stokes parameters and the sign is determined by the sign of the total quantity under the square root. All polarized spectra presented herein are displayed in this manner. Note, however, that at low S/N, $P_{\rm db}$ is also not a reliable function, as it has a peculiar probability distribution \citep{Miller1988}. Thus, for extracting statistically reliable values of polarization within a particular waveband, it is best to bin the calibrated $Q$ and $U$ Stokes spectra separately over the wavelength range of interest before calculating $P$ and $\theta$; all quoted and tabulated values in this paper were determined in this manner, although spectra are displayed as $P_{\rm db}$, so may exhibit slight offsets from our quoted values.

Table~1 lists the observation dates, instruments used, and the average $V$-band polarization ($P_V$) and position angle ($\theta$) derived from the polarized spectra. We first binned $Q$ and $U$ separately in the $V$-band wavelength range 5050--5950\,{\AA} and calculated $P_V$ and $\theta$ from there. Most of the observations with the Kuiper and Bok telescopes at Steward Observatory were conducted on consecutive nights. In cases where there was no statistically significant change from night to night, we averaged those data to increase the S/N.  

\subsection{Weak interstellar polarization}
All available data indicate that the interstellar extinction toward SN\,2017hcc is low. From high-resolution spectra of the interstellar Na\,{\sc i}\,D absorption, \citet{Smith2020} estimated a host-galaxy reddening of $E(B-V)\la0.016$\,mag, or $A_V\la0.05$\,mag, which is lower than the estimated Milky Way line-of-sight reddening of $E(B-V)=0.0285$\,mag. Considering the so-called Serkowski relation, whereby $P_{\rm ISM}<9\times E(B-V)$, the inferred Milky Way line-of-sight reddening of $E(B-V)=0.0285$\,mag \citep{Prieto2017}, combined with a small host reddening of $E(B-V)\la0.016$\,mag \citep{Smith2020}, implies interstellar polarization (ISP) $<0.4$\%; a maximum value of 0.4\% would only occur if the position angles of the host and Milky Way polarization were aligned and constructively interfered. Alternatively, if these separate components were perfectly misaligned and interfered destructively, then the ISP could be $<0.1$\%. It should be noted, however, that this analysis requires that the interstellar dust in both the Milky Way and the SN host galaxy polarize photons according to the Serkowski law.

We re-examine the ISP in Section 3.2.4 to further demonstrate its low value, but state from the outset that,  given the large degree of polarization measured for the SN, the small degree of possible ISP is relatively unsubstantial and does not influence our physical interpretation. We thus make no attempt to remove the ISP from the data. 

\subsection{Spectropolarimetric evolution of SN 2017hcc}
\begin{figure}
\includegraphics[width=3.2in]{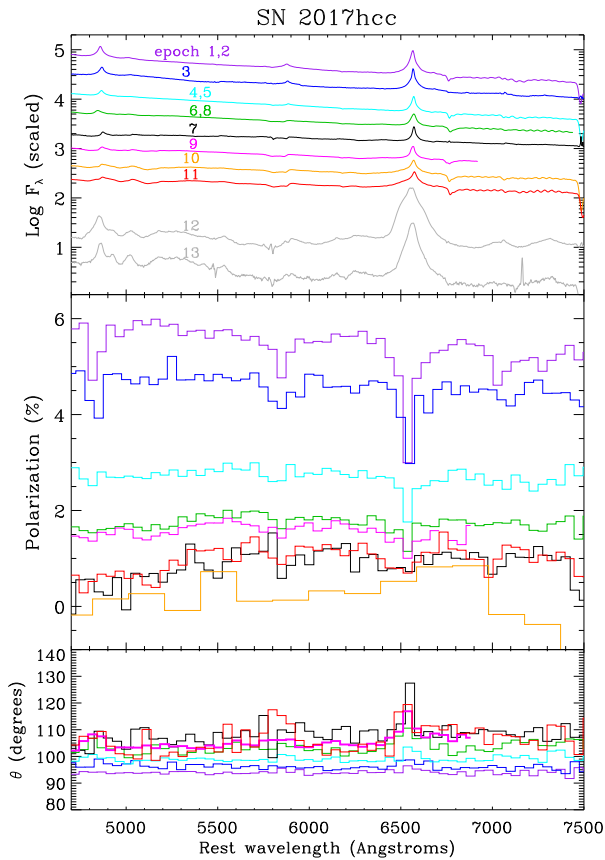}
\caption{Flux, polarization $P_{\rm db}$, and position angle $\theta$ for SN\,2017hcc. The data are averaged into epochs where multiple collections were on consecutive nights of a given observing campaign (see Column~1 of Table~1). The data for $P$ and $\theta$ are binned to 50\,{\AA} resolution, except for the day 103 data (200\,{\AA}) which were of lower quality. Error bars are omitted for clarity. For the latest two epochs on days 291 and 391 only the total-flux spectra are shown, as $P$ and $\theta$ data for those epochs have low S/N and provide only upper limits on the integrated polarization. The low-S/N $\theta$ spectrum for day 103 is also omitted for clarity. }
\label{fig:pol_seq}
\end{figure}

\begin{figure}
\includegraphics[width=3.2in]{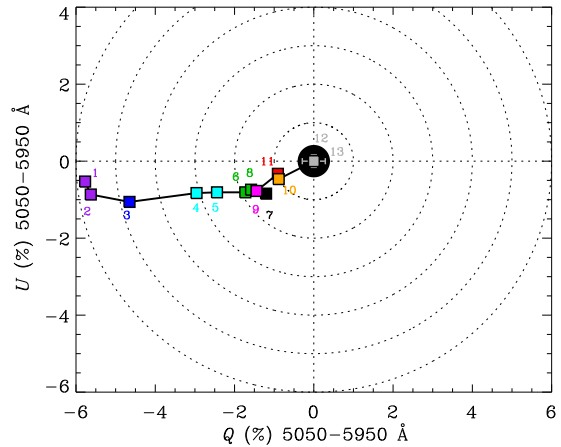}
\caption{Evolution of the $V$ band in the $Q$--$U$ plane at epochs 1--13, corresponding to days 16--391 days post-explosion. Interstellar polarization is constrained to $< 0.4$\%, illustrated by the black dot near the origin. The colors of the symbols correspond to those epochs used to create the averaged spectra displayed in Figure 1 with the same color display. In most cases the error bars are smaller than the symbols.}
\label{fig:qu}
\end{figure}

Figure~\ref{fig:pol_seq} shows the total-flux spectra, fractional polarization ($P$), and position angle ($\theta$) of SN\,2017hcc for epochs between days 16 and 109, and two late-time epochs on days 291 and 391.  Figure~\ref{fig:qu} displays the temporal evolution of the integrated $V$-band polarization (5050--5950\,\AA) in the $Q$--$U$ plane, which we regard as a continuum sample. Below we address the spectropolarimetric evolution of the continuum, prominent emission lines, and position angle.

\subsubsection{Continuum polarization}
\begin{figure*}
\includegraphics[width=7.1in]{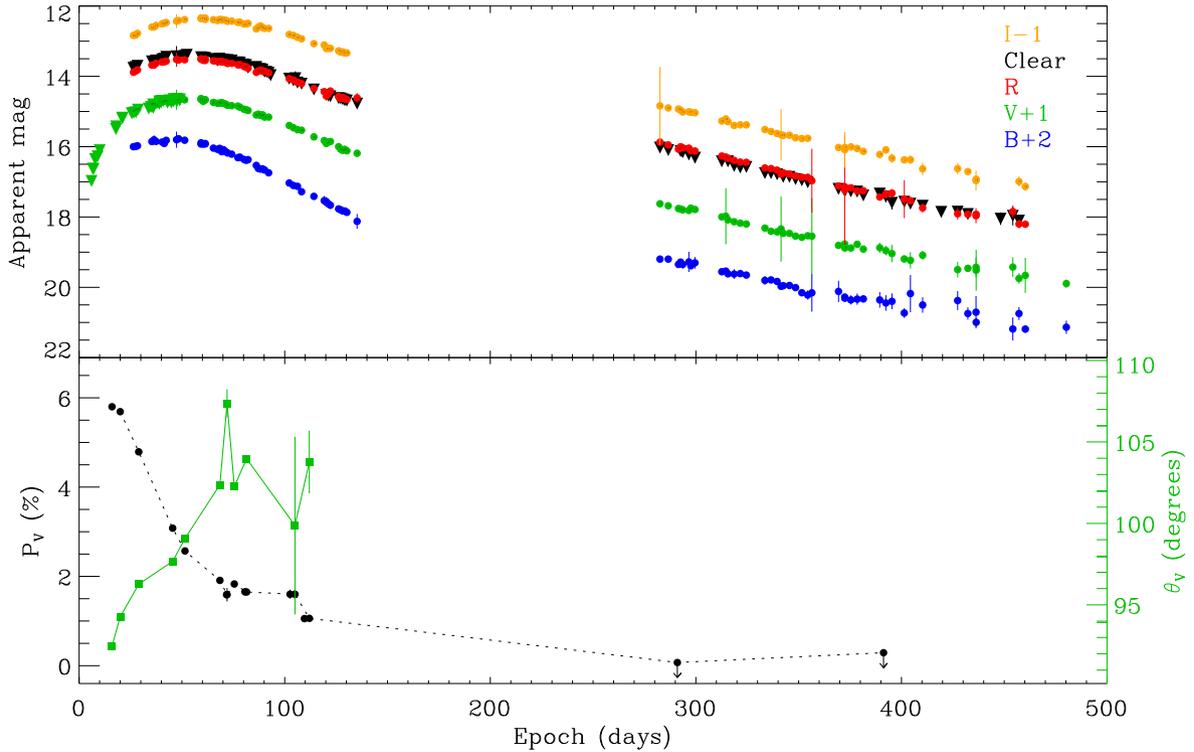}
\caption{{\it Upper panel:} {\it BVRI} and {\it Clear}-band light curve from KAIT and the Nickel telescope (filled circles), and early $V$-band data from \citet{Prieto2017} shown as inverted triangles. A small sample of KAIT data points in the range 300--400\,days are excluded owing to their large uncertainties. {\it Lower panel:} Polarization and $\theta$ averaged over the 5050--5950\,{\AA} $V$-band region, with the exception of day 291, which was averaged over the continuum region 6400--6700\,{\AA}.}
\label{fig:pol_lc}
\end{figure*}

At the earliest observed epoch 16 days after explosion, SN\,2017hcc exhibits a strongly polarized continuum with $P_V=5.8$\%. The continuum slope rises toward blue wavelengths by nearly 1\% between 7500\,\AA\ and 4700\,\AA, reaching nearly 6\% at the blue end. The wavelength dependence is observed through day 29, at which point $P_V$ has declined to 4.8\%. By day 45, however, the wavelength dependence has disappeared. Interestingly, the steep decline in $P$ and the change in wavelength dependence occurs while the SN flux is still climbing to peak brightness, as illustrated in Figure~\ref{fig:pol_lc}. By day 51, $P_V$ has dropped to 2.6\%, while the SN flux is still climbing to its $V$-band peak one week later on approximately MJD 58087 (day $\sim 59$). By day 68 the SN has passed peak flux and the polarization has dropped to 1.9\%. At this point in time the decline rate of the polarization has slowed. By day 103, $P_V$ drops more slowly to 1.6\%, and bottoms out at $\sim 1.1$\% on day 110. Very late epochs of spectropolarimetry obtained on days 291 and 391 provide only upper limits; these data have large uncertainties, as the source was faint and (on day 291) observed in morning twilight.  Nevertheless, the day 291 data indicate a very low continuum polarization level below 0.1\%, which is physically significant as discussed later.

\subsubsection{Continuum position angle}
Overall, $\theta$ is remarkably constant throughout the observed evolution of SN\,2017hcc, shifting by $< 10^\circ$ during the steep polarization decline between days 16 and 68. The $Q$--$U$ track  shifts direction slightly by a few degrees as the polarization declines further and the source approaches the origin of the $Q$--$U$ plane, but overall, deviations in $\theta$ are always $< 15^\circ$, and mostly $\lesssim10^\circ$.  For the very late epochs on days 291 and 391, $\theta$ has no meaning, and the source is consistent with zero intrinsic polarization, which is expected from the inferred small degree of ISP. 

\begin{figure}
\includegraphics[width=3.2in]{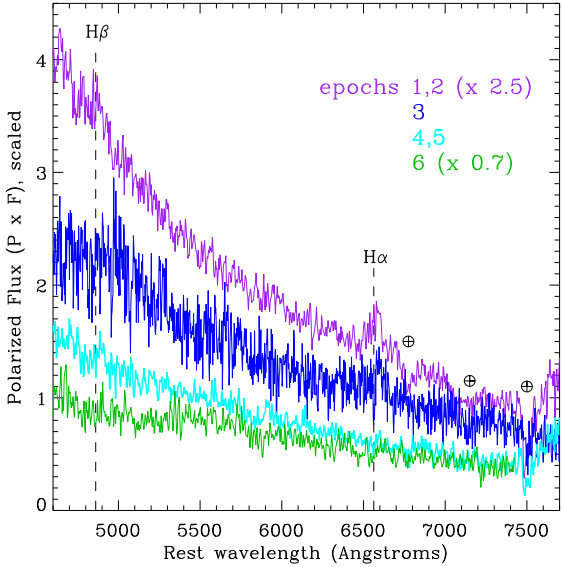}
\caption{Polarized flux (scaled vertically for clarity) for epochs 1--6. The colors correspond to those in Figures 1 and 2. The wavelengths of the H$\alpha$ and H$\beta$ emission are marked by dashed vertical lines. Wavelengths of telluric absorption features are illustrated by Earth symbols (circled ``plus" signs).}
\label{fig:pflux}
\end{figure}

\subsubsection{Line polarization}
During the first few months after explosion the spectrum exhibits strongly depolarized cores for the narrow Balmer and He\,{\sc i} emission lines. The core of H$\beta$ is less depolarized than that of H$\alpha$, possibly the result of the stronger underlying polarized continuum at those shorter wavelengths. We note that the spectral resolution of our data limits our sensitivity to the actual polarization minimum of the narrow line cores, as the strong  polarization of the surrounding continuum, and possibly the wings of the Lorentzian emission profile, probably blurs and fills in some of the line. 

The relatively narrow emission in SNe~IIn originates in the pre-shock gas of the CSM, which is illuminated by ultraviolet (UV) photons from the underlying shock interaction region \citep{smith17}; as this narrow-line region from excited CSM gas extends to radii outside the electron-scattering zone, it contributes unpolarized photons to the source, which leads to depolarization of the line core. This line depolarization of H$\alpha$ and H$\beta$ disappears by day 72. Interestingly, day 75 is when \citet{Smith2020} began to see the direct emission from the fast SN ejecta in the form of broad He\,{\i} P~Cygni absorption. Up until this point the strongly depolarized lines exhibited no significant  change across the line. Then on day 68 we start to see a PA change develop across the line as the depolarization fills in, becoming most prominent on day 72, exhibiting a $\sim 20^{\circ}$ PA excursion across H$\alpha$. The high-resolution spectra of \citet{Smith2020} also showed that at those later times the narrow emission from the pre-shock CSM weakened, while the P~Cygni profile developed a prominent absorption component that was actually stronger than the corresponding narrow emission.  The data from \citet{Smith2020} therefore might explain why the depolarization disappears in our lower-resolution polarized spectra, which does not resolve the P~Cygni profile seen by their high-resolution spectra --- that is, because narrow emission cannot cause net depolarization if the narrow absorption component of the profile becomes stronger. 

\subsubsection{Polarized Flux ($P \times F$)}
The polarized flux, given by the product of the total-flux spectra and fraction polarization, is shown in  Figure~\ref{fig:pflux}. Epochs 1--2 and 4--5, which correspond to days 16--21 and 45--52 (respectively), have been averaged to increase the S/N. The data have also been scaled vertically for visual clarity. The motivation for showing this plot is to determine whether any of the prominent emission lines, which appear strongly depolarized when displayed as fractional polarization (see Figure~\ref{fig:pol_seq}), exhibit evolving features in polarized flux. Interestingly, H$\alpha$ exhibits significant polarized flux for the first three epochs, when the overall polarization of the source is extremely strong and the fractional polarization of the continuum exhibits a blue slope (see Figure~\ref{fig:pol_seq}). There is also an H$\beta$ feature for epochs 1 and 2, which becomes indistinguishable from noise by epoch 3. The PA is stable across H$\alpha$, within a few degrees, as illustrated by Figure~\ref{fig:pol_seq}. The implications of our detection of polarized flux for H$\alpha$ during the early epochs will be discussed in Section 4.

\subsubsection{ISP revisited}
We further demonstrate the relatively low value of the ISP using our late-time data from day 391 (epoch 13). The highest S/N portion of the flux spectrum during that epoch is the region of the strong H$\alpha$ emission feature, for which  $Q$ and $U$ are both a null detection. Assuming the line emission is intrinsically unpolarized during this late optically-thin phase, we can calculate an upper limit ISP from the noise in the polarized spectrum across the line. The standard deviation of $Q$ and $U$ over the full-width at half-maximum intensity of the line, between 6650\,\AA\ and 6700\,\AA, gives $Q<0.37$\% and $U<0.46$\%, implying $P<0.59$\% (1$\sigma$). This limit is not as constraining as the limit implied by the low $E(B-V)$ color index in Section 3.1, because at this late phase the SN was relatively faint at $V\approx17$\,mag and the S/N is low, but the two estimates are nonetheless mutually consistent and further support our claim that ISP does not significantly impact our interpretation.

\section{Discussion}
Continuum polarization in SNe has generally been attributed to electron scattering in the photosphere, which can generate a net continuum polarization if the outflow geometry and projection on the sky are aspherical. For interacting SNe~IIn, as the SN shock wave plows through CSM, a dense shell of ionized material forms in the wake, creating a pseudophotosphere outside of the SN outflow photosphere that traces the geometry of the SN/CSM interface. At early times, a photoionized precursor often places the continuum photosphere upstream in the unshocked CSM itself.  Net continuum polarization can therefore arise from an aspherical distribution of CSM, regardless of the underlying SN outflow geometry. This is probably why interacting SNe are more likely to be polarized as a class, because CSM distributions around evolved massive stars can have complex and aspherical geometries \citep{Wachter2010, Clark05, smith01,smith07}. Meanwhile, pre-shock CSM gas surrounding this region becomes photoionized by UV photons and/or X-rays, and the resulting emission lines contribute unpolarized photons to the source, reducing the polarization fraction at those discrete wavelengths. 

\begin{figure}
\includegraphics[width=3.1in]{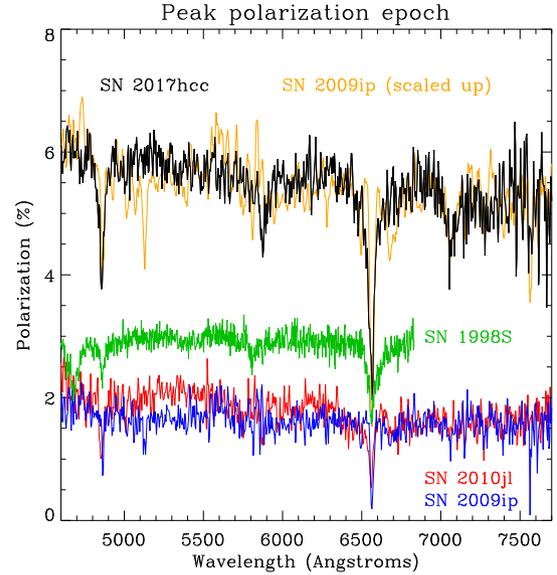}
\caption{Polarization spectra at observed peak polarization epoch for SN\,2017hcc, compared with SN\,2009ip \citep{Mauerhan2014}, SN\,2010jl (Lick/Kast epoch from Williams et al., in prep), and SN\,1998S \citep{Leonard2000}. A version of the SN\,2009ip polarized spectrum, scaled up by a factor of 3.35 to match the strength of SN\,2017hcc and smoothed by factor of 4, is shown to illustrate the overall similarities and slope.}
\label{fig:sn_comp}
\end{figure}

The spectropolarimetric data on SN\,2017hcc appear roughly consistent with this picture, exhibiting a strongly polarized continuum and partially depolarized emission-line cores. However, electron scattering is wavelength-independent \citep{Miller90,Miller91} and, yet, during the first month we detect a significant slope in the continuum polarization, rising $\sim 1$\% toward blue wavelengths in the range 4700--7500\,{\AA}, as shown in Figure~\ref{fig:sn_comp}. Dust scattering has higher polarization efficiency than electron scattering, in addition to a wavelength dependence (e.g., see \citealt{Kawabata2000} and references therein), which would naturally explain the unusually strong continuum polarization and its blue slope during the first month. Moreover, the net polarized flux ($P\times F$) for H$\alpha$ and H$\beta$, shown in Figure~\ref{fig:pflux}), is also consistent with dust scattering of the SN~IIn spectrum during the first month, as such line emission should be intrinsically unpolarized.  Analyses of IR spectra of SN\,2017hcc indicate that the CSM around the progenitor contained dust that must have been pre-existing \citep{Chandra2022, moran2023, Smith2020}, so some scattering of SN light by this dust is to be expected.  Similarly, IR spectra of SN~2009ip also revealed dust in the immediate pre-SN environment \citep{smith13}, and as we note below, SN~2009ip showed an early-time blue excess in the polarization similar to that of SN2017hcc.

The remarkable drop in fractional polarization from $\sim6$\% to $\sim3$\% during the first 45\,days appears to be anticorrelated with the SN light curve as it climbs to peak brightness. This drop in polarization is also associated with the disappearance of both the blue-rising slope of the polarized continuum and the polarized flux of the Balmer lines. This implies a transition in the relative contributions of dust versus free electrons to the scattering media and polarized flux. There are several potential causes for this transition. 

First, the scattering transition might be attributable to the structure and composition of the CSM. Multiwavelength analysis of SN\,2017hcc \citep{Smith2020,Chandra2022} indicates that the progenitor experienced a brief period of strongly enhanced mass loss in the decade prior to explosion, creating a dense inner CSM envelope that scattered the SN light as the shock plowed through it during the first days to months after explosion. \citep{Smith2020} suggest that a significant fraction of the $\sim$10 $M_{\odot}$ CSM in this inner envelope was likely concentrated in an equatorial region.   The interaction timescale for the inner CSM appears consistent with the polarization decline timescale and scattering transition. If the dense inner envelope inferred was relatively dust-rich, then this might explain the drop in strength and a change in the mode of polarization once this inner region was overtaken by the shock.

One should also consider the possibility that the UV luminosity of the superluminous SN heated and sublimated dust in the CSM, and whether this process could affect the relative contribution of electron scattering versus dust scattering as the SN climbed to peak. Interestingly, some recent theoretical investigations of radiative disruption of dust grains in an SN environment suggest that timescales of $\sim1$--2 months are significant for grain evaporation by superluminous SNe \citep{Hoang2019}. Incidentally, \citet{moran2023} reported that the blackbody temperature associated with the IR excess of SN\,2017hcc inferred on day 110, for example, is 1900\,K, which is above the evaporation temperature for grains of both graphite \citep{stritzinger2012} and silicates \citep{laor_draine1993,Gall2014}. Therefore, it seems plausible that dust in the inner  CSM could have scattered photons more efficiently during the first month or so before the luminosity reached its peak and the dust was broken down by the UV radiation field. 

 Whatever scenario governed the spectropolarimetric transition, the relative constancy of the position angle between the purported dust-scattering phase ($<45$\,days) and electron-scattering phase ($>45$\,days) implies that the scattering material of these two components are part of the same geometric structure. If, instead, the light echoes came from a mottled or irregular distribution of dust clumps or shells, then we might expect more pronounced fluctuations in PA between transition and over the course of the SN evolution. Rather, as illustrated by Figure~\ref{fig:qu}, the PA of the source varies little during the entire spectropolarimetric evolution and follows a relatively straight path with time toward the origin in the $Q$--$U$ plane. Therefore it appears that as the SN shock propagated outward it continued to interact with the same CSM geometry that echoed the SN light at early times, before it was either engulfed or the dust sublimated. Therefore, the electron-scattering pseudophotosphere resulting from that interaction produces the same projected scattering geometry on the sky as the light echo, and results in a relatively stable PA as the SN transitions between these two scattering phases. Moreover, the dust scattering and electron scattering components having the same geometry could also explain the extreme record-breaking polarization, since the integrated electric vectors of both scattering components would interfere constructively and amplify the net polarization signal. This consistent well-ordered geometry for all significant sources of polarization in the mix implies the presence of a continuous CSM structure with a global geometry that is preserved at least out to radii the SN has reached during the course of our spectropolarimetric coverage.  
 
It should also be mentioned that the very low or undetectable polarization at late times, in contrast to the high polarization at early times, has implications for the origin of the intermediate-width components in the spectral lines.  At early times, as typically seen in SNe~IIn, SN\,2017hcc showed strong Balmer emission lines with symmetric Lorentzian-shaped wings.  These Lorentzian profiles are thought to be caused by electron scattering of the narrow-line emission \citep{Chugai2001}. Data from \citet{Smith2020} showed that at later epochs, especially after day 200 in SN\,2017hcc, the intermediate-width components no longer showed Lorentzian shapes and instead transitioned to composite shapes (both broad and intermediate-width) that were marginally asymmetric and skewed to the blue. Although our spectral data do not have sufficient resolution to examine the polarized line profile in detail, the lack of detectable polarization at these later epochs is consistent with the interpretation of \citet{Smith2020} that the narrow lines were no longer broadened by electron scattering, but instead, their widths indicate that Doppler broadening from expansion of the post-shock gas and SN ejecta was responsible for the line profile.

A transition from a polarized source initially dominated by dust scattering to later becoming dominated by electron scattering might also have been exhibited by SN\,2009ip, where a relatively rapid rise and drop in polarization during Oct. 2012 was also associated with the appearance of a blue-rising continuum polarization \citep{Mauerhan2014}. A re-examination of the spectropolarimetry data for SN\,2009ip from \citet{Mauerhan2014}, not shown here, also revealed that net polarized flux ($P\times F$) for the H$\alpha$ line was also only seen during the rapid polarization spike in Oct. 2012; we suspect that, as in the case of SN\,2017hcc, an extended distribution of dusty CSM echoed the SN~IIn spectrum during that phase, including the narrow H$\alpha$ from the CSM interaction occurring within. Aside from the much higher polarization strength of SN\,2017hcc, the overall structure of the polarized spectrum actually appears quite similar to those of the Type IIn SN\,2009ip \citep{Mauerhan2014} and SN\,2010jl (Williams et al., in prep.). Figure~\ref{fig:sn_comp} compares each of these SNe at their highest observed polarization epoch; SN\,1998S data from \citet{Leonard2000} are also shown for comparison. Interestingly, if one simply multiplies the polarized spectrum of SN\,2009ip during its observed polarization peak by a factor of 3.35, the polarized spectrum matches that of SN\,2017hcc remarkably well, including the dust-scattering-induced slope of the polarized continuum and the depolarization fraction of the emission lines. 

The extreme record-breaking polarization of SN\,2017hcc likely indicates that the SN/CSM is intrinsically more aspherical than that of SN\,2009ip and other SNe observed with polarimetry thus far, and/or viewed at an orientation angle that enhances the projected asphericity on the plane of the sky.  In many previous studies, SN asphericity has been quantitatively parameterized by the axis ratio of a simple ellipse \citep{Hoeflich1991}, but it is only appropriate to quantify asphericity in this way for the case of pure electron scattering. Nonetheless, it remains true that a higher degree of projected asphericity will result in a higher degree of polarization, regardless of whether the polarized flux is dominated by dust scattering or electron scattering. It is also possible that SN\,2017hcc's stronger polarization results in part from the fact that it had a relatively large mass of CSM with more light-scattering dust.

In conclusion, it is clear that multicomponent scattering models incorporating both dust and electron scattering are needed to quantify SN/CSM geometry based on polarization data with a higher degree of confidence. As the sample of polarized SNe continues to increase, there will be more opportunities to search for statistical correlations between polarization level and other physical parameters, which will help elucidate the physics and geometries involved. Interpretive challenges notwithstanding, it is abundantly clear that the collective spectropolarimetric data on SNe~IIn imply the commonality of aspherical distributions of CSM for SN~IIn progenitors, and this prevalence seemingly dovetails with the hypothesis that binary interactions between highly evolved massive stars, whose envelopes expand during their final nuclear-burning stages and possibly come into contact with one another,  play an important role in shaping the CSM SN~IIn progenitor systems \citep{Smith2014,sa14}. 


\section*{Acknowledgements}
\scriptsize 
Some of the observations reported herein were obtained at the MMT Observatory, a joint facility of the University of Arizona and the Smithsonian Institution. 
A major upgrade of the   Kast spectrograph on the Shane 3\,m telescope at Lick Observatory was     made possible through generous gifts from William and Marina Kast as      well as the Heising-Simons Foundation. 
KAIT and its ongoing operation were made possible by donations from Sun Microsystems, Inc., the Hewlett-Packard Company, AutoScope 
Corporation, Lick Observatory, the U.S. NSF, the University of
California, the Sylvia \& Jim Katzman Foundation, and the TABASGO 
Foundation.
We thank the staff at Lick and Steward Observatories for their excellent observing support. Research at Lick Observatory is partially supported by a generous gift from Google. Help obtaining images with the 1\,m Nickel telescope was provided by U.C. Berkeley graduate student Sergiy Vasylyev, as well as undergraduate students
Sanyum Channa,
Romain Hardy,
Julia Hestenes,
Edward Falcon,
Nachiket Girish,
Andrew Hoffman, 
Evelyn Liu, 
Emily Ma,
Yukei Murakami,
Timothy Ross,
Jackson Sipple, 
James Sunseri,
Kevin Tang,
Jeremy Wayland,
Abel Yagubyan,
and
Sameen Yunus. D.C.L. acknowledges support from NSF grants AST-1210311 and AST-2010001, under which part of this research was carried out. A.V.F.'s supernova group at U.C. Berkeley has received financial assistance from the Christopher R. Redlich Fund, Alan Eustace, Kathleen and Frank Wood, and many other individual donors.
\scriptsize
\bibliographystyle{mnras}
\bibliography{sn2017hcc_specpol}
\label{lastpage}

\section{Appendix}

\begin{table*}
\centering
\caption{KAIT and Nickel photometry of SN\,2017hcc; $t_0$ is the adopted explosion date, MJD 58027.9}
\begin{tabular}{|l|r|c|c|c|c|c|c|}
\hline
MJD & $t-t_{0}$ (days) & $B$ & $V$ & $R$ & $I$ & $Clear$ & Tel. \\
\hline
58054.3086  &   26.4  & 14.00 (0.02) & 13.98 (0.01) & 13.88 (0.01) & 13.84 (0.01) & 13.76 (0.01) &    KAIT \\
58055.3086  &   27.4  & 13.99 (0.02) & 13.96 (0.01) & 13.85 (0.01) & 13.83 (0.01) & 13.74 (0.01) &    KAIT \\
58056.2969  &   28.4  & 13.98 (0.02) & 13.93 (0.01) & 13.82 (0.01) & 13.78 (0.01) & 13.70 (0.01) &    KAIT \\
58063.2734  &   35.4  & 13.85 (0.02) & 13.78 (0.01) & 13.68 (0.01) & 13.60 (0.01) & 13.57 (0.01) &    KAIT \\
58064.3359  &   36.4  & 13.81 (0.13) & 13.79 (0.02) & 13.68 (0.01) & 13.60 (0.01) & 13.61 (0.01) &    KAIT \\
58065.3008  &   37.4  & 13.85 (0.10) & 13.81 (0.06) & 13.63 (0.08) & 13.55 (0.05) & 13.55 (0.02) &    KAIT \\
58068.3125  &   40.4  & 13.87 (0.02) & 13.72 (0.01) & 13.60 (0.01) & 13.51 (0.01) & 13.51 (0.01) &    KAIT \\
58069.3164  &   41.4  & 13.91 (0.02) & 13.70 (0.01) & 13.59 (0.01) & 13.49 (0.01) & 13.50 (0.01) &    KAIT \\
58070.2578  &   42.4  & 13.82 (0.02) & 13.69 (0.01) & 13.57 (0.01) & 13.47 (0.01) & 13.45 (0.02) &    KAIT \\
58075.2969  &   47.4  & 13.80 (0.23) & 13.66 (0.28) & 13.52 (0.20) & 13.43 (0.20) & 13.43 (0.29) &    KAIT \\
58076.2891  &   48.4  & 13.77 (0.03) & 13.68 (0.01) & 13.54 (0.01) & 13.41 (0.02) & 13.46 (0.01) &    KAIT \\
58079.2617  &   51.4  & 13.82 (0.02) & 13.67 (0.01) & 13.53 (0.01) & 13.39 (0.01) & 13.42 (0.01) &    KAIT \\
58080.2695  &   52.4  &              &              &              &              & 13.39 (0.14) &    KAIT \\
58087.2305  &   59.3  & 13.89 (0.04) & 13.63 (0.02) & 13.51 (0.02) & 13.34 (0.02) & 13.46 (0.02) &    KAIT \\
58088.2422  &   60.3  & 13.93 (0.04) & 13.71 (0.01) & 13.54 (0.01) & 13.36 (0.01) & 13.47 (0.01) &    KAIT \\
58089.2266  &   61.3  & 13.91 (0.05) & 13.66 (0.01) & 13.55 (0.01) & 13.38 (0.02) & 13.48 (0.01) &    KAIT \\
58093.2266  &   65.3  & 14.04 (0.04) & 13.74 (0.01) & 13.56 (0.01) & 13.39 (0.01) & 13.51 (0.01) &    KAIT \\
58095.2461  &   67.3  & 14.08 (0.02) & 13.77 (0.01) & 13.58 (0.01) & 13.41 (0.01) & 13.55 (0.01) &    KAIT \\
58096.2109  &   68.3  & 14.05 (0.02) & 13.76 (0.01) & 13.57 (0.01) & 13.38 (0.01) & 13.50 (0.01) &    KAIT \\
58097.2227  &   69.3  & 14.08 (0.02) & 13.76 (0.01) & 13.58 (0.01) & 13.40 (0.01) & 13.52 (0.01) &    KAIT \\
58098.1523  &   70.3  & 14.09 (0.04) & 13.77 (0.01) & 13.59 (0.01) & 13.39 (0.01) & 13.52 (0.01) &    KAIT \\
58099.1680  &   71.3  & 14.14 (0.03) & 13.79 (0.01) & 13.60 (0.01) & 13.40 (0.01) & 13.54 (0.01) &    KAIT \\
58100.1016  &   72.2  & 14.18 (0.02) & 13.82 (0.02) & 13.63 (0.02) & 13.44 (0.03) & 13.54 (0.05) &    KAIT \\
58101.1680  &   73.3  &              &              &              &              & 13.56 (0.02) &    KAIT \\
58102.1172  &   74.2  & 14.21 (0.01) & 13.83 (0.01) & 13.63 (0.01) & 13.43 (0.01) & 13.55 (0.01) &    KAIT \\
58105.1367  &   77.2  & 14.31 (0.04) & 13.85 (0.02) & 13.67 (0.01) & 13.47 (0.02) & 13.60 (0.01) &    KAIT \\
58106.1289  &   78.2  & 14.30 (0.02) & 13.88 (0.01) & 13.66 (0.01) & 13.46 (0.01) & 13.61 (0.01) &    KAIT \\
58108.1172  &   80.2  & 14.36 (0.02) & 13.94 (0.01) & 13.72 (0.01) & 13.51 (0.01) & 13.66 (0.01) &    KAIT \\
58109.1211  &   81.2  & 14.39 (0.02) & 13.95 (0.01) & 13.73 (0.01) & 13.51 (0.01) & 13.68 (0.01) &    KAIT \\
58110.0938  &   82.2  & 14.37 (0.10) & 13.97 (0.04) & 13.77 (0.05) & 13.49 (0.08) & 13.75 (0.03) &    KAIT \\
58114.0977  &   86.2  & 14.53 (0.03) & 14.10 (0.02) & 13.89 (0.02) & 13.65 (0.03) & 13.73 (0.05) &    KAIT \\
58115.1133  &   87.2  & 14.62 (0.03) & 14.08 (0.02) & 13.84 (0.01) & 13.60 (0.02) & 13.79 (0.01) &    KAIT \\
58116.1680  &   88.3  & 14.64 (0.06) & 14.11 (0.02) & 13.83 (0.02) & 13.56 (0.02) & 13.79 (0.01) &    KAIT \\
58117.1094  &   89.2  & 14.63 (0.02) & 14.09 (0.01) & 13.84 (0.01) & 13.61 (0.01) & 13.81 (0.01) &    KAIT \\
58118.0938  &   90.2  & 14.66 (0.08) & 14.16 (0.04) & 13.85 (0.03) & 13.64 (0.02) & 13.87 (0.03) &    KAIT \\
58119.1094  &   91.2  &              &              &              &              & 13.88 (0.03) &    KAIT \\
58120.1055  &   92.2  & 14.74 (0.06) & 14.16 (0.03) & 13.90 (0.01) & 13.64 (0.02) & 13.90 (0.02) &    KAIT \\
58121.1016  &   93.2  &              &              &              &              & 13.97 (0.10) &    KAIT \\
58130.1133  &  102.2  & 15.03 (0.02) & 14.40 (0.01) & 14.07 (0.01) & 13.81 (0.01) & 14.07 (0.01) &    KAIT \\
58132.1211  &  104.2  & 15.10 (0.02) & 14.46 (0.01) & 14.10 (0.02) & 13.84 (0.02) & 14.11 (0.01) &    KAIT \\
58133.0977  &  105.2  &              &              &              &              & 14.05 (0.20) &    KAIT \\
58134.1172  &  106.2  & 15.13 (0.03) & 14.52 (0.02) & 14.16 (0.02) & 13.89 (0.02) & 14.18 (0.01) &    KAIT \\
58136.1172  &  108.2  & 15.28 (0.03) & 14.53 (0.02) & 14.21 (0.01) & 13.93 (0.02) & 14.21 (0.01) &    KAIT \\
58142.1055  &  114.2  & 15.41 (0.03) & 14.72 (0.01) & 14.35 (0.01) & 14.07 (0.02) & 14.38 (0.02) &    KAIT \\
58148.1250  &  120.2  & 15.56 (0.13) & 14.91 (0.08) & 14.56 (0.06) & 14.21 (0.06) & 14.53 (0.03) &    KAIT \\
58149.1211  &  121.2  & 15.63 (0.05) & 14.87 (0.02) & 14.49 (0.01) & 14.20 (0.01) & 14.55 (0.01) &    KAIT \\
58150.1289  &  122.2  & 15.66 (0.11) & 14.85 (0.07) & 14.42 (0.04) & 14.20 (0.04) & 14.57 (0.02) &    KAIT \\
58154.1328  &  126.2  & 15.77 (0.07) & 15.00 (0.02) & 14.58 (0.02) & 14.27 (0.02) & 14.62 (0.02) &    KAIT \\
58155.1250  &  127.2  & 15.80 (0.05) & 15.07 (0.02) & 14.63 (0.02) & 14.31 (0.02) & 14.63 (0.02) &    KAIT \\
58156.1328  &  128.2  & 15.83 (0.07) & 15.11 (0.03) & 14.60 (0.02) & 14.31 (0.02) & 14.65 (0.01) &    KAIT \\
58157.1172  &  129.2  & 15.83 (0.04) & 15.07 (0.02) & 14.63 (0.01) & 14.34 (0.02) & 14.68 (0.01) &    KAIT \\
58158.1328  &  130.2  & 15.87 (0.06) & 15.12 (0.02) & 14.66 (0.02) & 14.33 (0.02) & 14.69 (0.01) &    KAIT \\
58163.1211  &  135.2  & 16.12 (0.21) & 15.19 (0.10) & 14.61 (0.14) &              & 14.78 (0.03) &    KAIT \\
58310.4531  &  282.6  & 17.20 (0.04) & 16.62 (0.09) & 15.87 (0.04) & 15.84 (1.11) & 16.03 (0.02) &    KAIT \\
58314.4766  &  286.6  & 17.20 (0.03) & 16.68 (0.02) & 15.95 (0.02) & 15.90 (0.02) & 16.10 (0.02) &    KAIT \\
58320.4688  &  292.6  & 17.28 (0.05) & 16.78 (0.02) & 16.00 (0.02) & 15.96 (0.02) & 16.13 (0.01) &    KAIT \\
58321.5078  &  293.6  & 17.35 (0.12) & 16.80 (0.05) & 16.03 (0.03) & 16.01 (0.05) & 16.16 (0.04) &    KAIT \\
58324.4961  &  296.6  & 17.28 (0.29) & 16.81 (0.06) & 16.04 (0.03) & 16.01 (0.04) & 16.22 (0.03) &    KAIT \\
58327.4883  &  299.6  & 17.30 (0.17) & 16.79 (0.08) & 16.12 (0.05) & 16.03 (0.05) & 16.32 (0.05) &    KAIT \\
58340.5117  &  312.6  & 17.56 (0.08) & 17.00 (0.04) & 16.27 (0.03) & 16.27 (0.03) & 16.42 (0.03) &    KAIT \\
58343.5000  &  315.6  & 17.62 (0.05) & 17.08 (0.03) & 16.33 (0.03) & 16.29 (0.03) & 16.42 (0.03) &    KAIT \\
58346.5195  &  318.6  & 17.62 (0.14) & 17.14 (0.06) & 16.39 (0.04) & 16.40 (0.06) & 16.55 (0.03) &    KAIT \\
58349.4883  &  321.6  & 17.61 (0.05) & 17.18 (0.05) & 16.44 (0.02) & 16.38 (0.03) & 16.59 (0.03) &    KAIT \\

\end{tabular}
\end{table*}

\clearpage

\begin{table*}\ContinuedFloat
\caption{continued}
\begin{tabular}{|l|c|c|c|c|c|c|c|}
\hline
MJD & $t-t_{0}$ (days) & $B$ & $V$ & $R$ & $I$ & $Clear$ & Tel. \\
\hline
58352.4531  &  324.6  & 17.65 (0.06) & 17.20 (0.03) & 16.44 (0.02) & 16.38 (0.03) & 16.58 (0.02) &    KAIT \\
58361.4844  &  333.6  & 17.80 (0.12) & 17.32 (0.09) & 16.61 (0.04) & 16.51 (0.05) & 16.75 (0.05) &    KAIT \\
58364.4531  &  336.6  & 17.79 (0.08) & 17.41 (0.05) & 16.62 (0.03) & 16.57 (0.04) & 16.75 (0.04) &    KAIT \\
58367.4102  &  339.5  & 17.83 (0.05) & 17.42 (0.04) & 16.67 (0.03) & 16.62 (0.03) & 16.79 (0.05) &    KAIT \\
58370.4102  &  342.5  & 17.95 (0.06) & 17.46 (0.05) & 16.74 (0.04) & 16.68 (0.04) & 16.86 (0.05) &    KAIT \\
58373.4023  &  345.5  & 17.95 (0.07) & 17.47 (0.06) & 16.78 (0.05) & 16.67 (0.05) & 16.86 (0.02) &    KAIT \\
58376.3867  &  348.5  & 18.01 (0.06) & 17.54 (0.04) & 16.83 (0.03) & 16.75 (0.04) & 16.92 (0.03) &    KAIT \\
58379.4414  &  351.5  & 18.15 (0.08) & 17.58 (0.04) & 16.87 (0.03) & 16.77 (0.04) & 16.96 (0.04) &    KAIT \\
58382.3281  &  354.4  & 18.22 (0.13) & 17.53 (0.06) & 16.87 (0.04) & 16.76 (0.06) & 17.01 (0.06) &    KAIT \\
58397.3008  &  369.4  & 18.12 (0.30) & 17.81 (0.07) & 17.12 (0.05) & 17.02 (0.06) & 17.20 (0.05) &    KAIT \\
58400.3789  &  372.5  & 18.31 (0.09) & 17.88 (0.06) & 17.14 (0.04) & 17.02 (0.05) & 17.27 (0.06) &    KAIT \\
58403.3164  &  375.4  & 18.36 (0.13) & 17.88 (0.09) & 17.17 (0.07) & 17.01 (0.09) & 17.27 (0.10) &    KAIT \\
58406.3789  &  378.5  & 18.34 (0.16) & 17.78 (0.09) & 17.26 (0.06) & 17.05 (0.08) & 17.28 (0.11) &    KAIT \\
58409.3672  &  381.5  & 18.33 (0.10) & 17.91 (0.06) & 17.25 (0.05) & 17.13 (0.06) & 17.39 (0.04) &    KAIT \\
58417.3438  &  389.4  & 18.36 (0.22) & 17.87 (0.14) & 17.43 (0.09) & 17.22 (0.10) & 17.33 (0.11) &    KAIT \\
58420.3281  &  392.4  & 18.45 (0.22) & 17.95 (0.14) & 17.35 (0.13) & 17.09 (0.11) & 17.42 (0.15) &    KAIT \\
58423.2852  &  395.4  & 18.40 (0.22) & 18.04 (0.25) & 17.32 (0.12) & 17.33 (0.13) & 17.60 (0.19) &    KAIT \\
58429.2734  &  401.4  & 18.73 (0.14) & 18.19 (0.09) & 17.49 (0.54) & 17.37 (0.08) & 17.57 (0.06) &    KAIT \\
58432.2695  &  404.4  & 18.18 (0.53) & 18.23 (0.23) & 17.54 (0.09) & 17.37 (0.12) & 17.62 (0.09) &    KAIT \\
58438.2109  &  410.3  & 18.50 (0.22) & 18.09 (0.13) & 17.74 (0.13) & 17.63 (0.18) & 17.68 (0.10) &    KAIT \\
58447.2617  &  419.4  &              &              &              &              & 17.86 (0.10) &    KAIT \\
58455.2266  &  427.3  & 18.38 (0.27) & 18.50 (0.22) & 17.91 (0.16) & 17.62 (0.16) & 17.85 (0.09) &    KAIT \\
58460.2031  &  432.3  & 18.75 (0.17) & 18.46 (0.09) & 17.92 (0.07) & 17.71 (0.07) & 17.92 (0.09) &    KAIT \\
58464.1797  &  436.3  & 18.71 (0.46) & 18.43 (0.27) & 17.96 (0.21) & 17.96 (0.28) &              &    KAIT \\
58476.1289  &  448.2  &              &              &              &              & 18.05 (0.18) &    KAIT \\
58482.0938  &  454.2  & 19.18 (0.33) & 18.42 (0.28) & 17.84 (0.17) &              & 17.97 (0.27) &    KAIT \\
58485.0898  &  457.2  & 18.75 (0.18) & 18.75 (0.15) & 18.20 (0.13) & 17.99 (0.15) & 18.10 (0.14) &    KAIT \\
58087.1914  &   59.3  & 13.93 (0.01) & 13.66 (0.01) & 13.51 (0.01) & 13.36 (0.01) &              &    KAIT \\
58089.2227  &   61.3  & 13.92 (0.02) & 13.67 (0.01) & 13.51 (0.01) & 13.34 (0.01) &              &  Nickel \\
58097.2461  &   69.3  & 14.14 (0.04) & 13.75 (0.01) & 13.59 (0.01) & 13.39 (0.01) &              &  Nickel \\
58147.1133  &  119.2  & 15.54 (0.13) & 14.83 (0.08) & 14.44 (0.05) & 14.11 (0.07) &              &  Nickel \\
58319.4375  &  291.5  & 17.35 (0.05) & 16.76 (0.02) & 16.07 (0.03) & 15.93 (0.03) &              &  Nickel \\
58325.4727  &  297.6  & 17.39 (0.03) & 16.75 (0.02) & 16.12 (0.02) & 16.02 (0.02) &              &  Nickel \\
58342.4688  &  314.6  & 17.54 (0.08) & 16.98 (0.80) & 16.30 (0.08) & 16.21 (0.06) &              &  Nickel \\
58369.3477  &  341.4  & 17.97 (0.06) & 17.34 (0.93) & 16.78 (0.10) & 16.66 (0.73) &              &  Nickel \\
58384.3398  &  356.4  & 18.16 (0.54) & 17.55 (1.08) & 16.97 (0.91) &              &              &  Nickel \\
58400.3164  &  372.4  & 18.28 (0.04) & 17.77 (0.03) & 17.23 (1.55) & 17.09 (0.51) &              &  Nickel \\
58464.2383  &  436.3  & 19.00 (0.13) & 18.51 (0.58) & 17.92 (0.07) & 17.92 (0.08) &              &  Nickel \\
58488.1641  &  460.3  & 19.19 (0.10) & 18.66 (0.50) & 18.20 (0.09) & 18.13 (0.12) &              &  Nickel \\
58508.1055  &  480.2  & 19.14 (0.19) & 18.89 (0.11) &              &              &              &  Nickel \\
58735.3594  &  707.5  & 20.22 (0.14) & 20.03 (0.14) &              &              &              &  Nickel \\
58751.2812  &  723.4  & 20.47 (0.73) & 19.83 (0.23) & 19.89 (0.67) & 19.66 (0.31) &              &  Nickel \\
58759.2422  &  731.3  & 20.47 (0.23) & 20.04 (0.40) &              & 20.06 (0.35) &              &  Nickel \\
58779.2812  &  751.4  & 20.48 (0.19) & 20.06 (0.16) & 20.23 (0.41) &              &              &  Nickel \\
58788.1875  &  760.3  & 20.72 (0.12) & 20.22 (0.37) & 20.02 (0.78) & 20.13 (0.21) &              &  Nickel \\
58789.2422  &  761.3  & 20.49 (0.11) & 20.29 (0.43) & 19.99 (0.21) & 19.70 (0.28) &              &  Nickel \\
58792.2031  &  764.3  & 20.51 (0.67) & 20.27 (0.49) & 19.88 (0.55) & 20.10 (0.20) &              &  Nickel \\
58805.2109  &  777.3  & 20.52 (0.48) & 20.38 (0.14) &              & 19.98 (0.16) &              &  Nickel \\
\end{tabular}
\end{table*}

\end{document}